\documentclass[aps,prl,floats,twocolumn,showpacs,superscriptaddress]
              {revtex4-1}      

\usepackage[dvips]{graphicx} 
\usepackage{epsfig}
\usepackage{subfigure}
\usepackage{amsmath}
\usepackage{epsfig}
\usepackage{color}

\newcommand{\f}{\hat{a}}
\newcommand{\fd}{\hat{a}^{\dagger}}
\newcommand{\fo}{\hat{c}}
\newcommand{\fod}{\hat{c}^{\dagger}}
\newcommand{\s}{\hat{S}}
\newcommand{\bj}{\bf j}

\begin{document}

\title{$XY\!Z$ Quantum Heisenberg Models with $p$-Orbital
  bosons}  
\author{Fernanda Pinheiro} 
\email{fep@fysik.su.se}  
\affiliation{Department of Physics, Stockholm University, Se-106 91
  Stockholm, Sweden} 
\affiliation{NORDITA,
KTH Royal Institute of Technology and Stockholm University, Se-106 91
Stockholm, Sweden}  
\author{Georg M. Bruun}
\affiliation{Department of Physics and Astronomy, University of
  Aarhus, DK-8000 Aarhus C, Denmark} 
\author{Jani-Petri Martikainen}
\affiliation{COMP Center of Excellence, Department of Applied Physics,
  Aalto University, Fi-00076, Aalto, Finland} 
\author{Jonas Larson}
\affiliation{Department of Physics, Stockholm University, Se-106 91
  Stockholm, Sweden} 
  \date{\today}

\begin{abstract}
We demonstrate how the spin-$1/2$ $XY\!Z$ quantum Heisenberg model
can be realized  with  bosonic atoms loaded in the $p$ band of an optical
lattice in the Mott regime. The combination of Bose statistics and the
symmetry  
of the $p$-orbital wave functions leads to a non-integrable  
Heisenberg model with  anti-ferromagnetic couplings. Moreover, the sign
and relative strength of the couplings characterizing the model are
shown to be experimentally tunable. We display the rich phase diagram
in the one dimensional case, and discuss finite size effects relevant
for trapped systems. Finally, experimental issues related to
preparation, manipulation, detection, and imperfections are
considered.  
\end{abstract}
\pacs{03.75.Lm, 67.85.Hj, 05.30.Rt}
\maketitle

{\it Introduction.}-- Powerful tools developed recently to
unravel the physics of many-body
quantum systems offer an exciting new platform for understanding
quantum magnetism. It is now possible to engineer
different systems in the lab that mimic the physics of theoretically
challenging spin models~\cite{maciek}, thereby performing ``quantum
simulations''~\cite{qs}. Along these lines,
systems of trapped ions and of polar molecules are promising
candidates. Trapped ions, for example, have already been
employed to simulate both small~\cite{trap1} and large~\cite{trap2}
numbers of spins. In these setups, however, sustaining control over
the parameters becomes very difficult as the system size increases.
Furthermore, due to trapping potentials
realizations are limited to chains with up to 25 spins. 
It is also very difficult to construct paradigmatic spin models with
short range interactions using systems of trapped ions.
Similar limitations appear when using polar molecules, where 
the effective spin interactions~\cite{polar1,polar2} are obtained
from the intrinsic dipole-dipole interactions.
Due to the character of the dipolar interaction, these systems give
rise to emergent models that are inherently long range and the
resulting couplings usually feature spatial anisotropies.     

Short range spin models can instead be realized with
 cold atoms in optical lattices~\cite{maciek}. A bosonic system in a
 tilted lattice has recently been used to simulate the phase
transition in a 1D Ising model~\cite{spin1}. Fermionic atoms were  
employed to study dynamical properties of quantum magnetism for  
spin systems~\cite{spin4,spin2}. This idea, first introduced in
Ref.~\cite{spin3}, has also 
been applied to other configurations, and simulation of different
types of spin models have been proposed~\cite{spinmodels}. However,
due to the character of the atomic $s$-wave scattering among the
different Zeeman levels, such mappings usually yield effective spin
models supporting continuous symmetries like the $X\!X\!Z$ model.
But as the main goal
of a quantum simulator is to realize systems that cannot be tackled
via analytical and/or numerical approaches, it is important to explore
alternative scenarios that yield low symmetry spin models with
anisotropic couplings and external fields.

In this paper we propose such a scenario by demonstrating 
that bosonic atoms in the first excited band ($p$ band) of a
two-dimensional (2D) optical lattice can realize the spin-$1/2$
$XY\!Z$ quantum Heisenberg model in an external field. Systems of cold 
atoms in excited bands feature an additional orbital degree of
freedom~\cite{girvin} that gives rise to novel 
physical properties~\cite{maciekliu}, which include 
supersolids~\cite{ss} and other types of novel phases~\cite{novphas}, 
unconventional condensation~\cite{nonzero}, and
frustration~\cite{frust}. 
Also a condensate with a complex
order parameter was recently observed experimentally~\cite{TMuller,complex}.
The dynamics of bosons in the $p$ band include anisotropic
tunneling and orbital changing interactions, where
two atoms in one orbital state  scatter into two atoms in a different
orbital state.   
This is the key mechanism leading to 
the anisotropy of the effective spin model obtained here: These
processes reduce the continuous $U(1)$ symmetry characteristic of the
$X\!X\!Z$ model, which would effectively describe fermions in the $p$
band~\cite{ref23}, into a set of discrete
$Z_2$ symmetries characteristic of the $XY\!Z$ 
model. In addition, due to the anomalous
$p$-band dispersions the couplings of the resulting spin model can
favor for anti-ferromagnetic order even in the bosonic case.

We also demonstrate how further control of
both the strength and sign of the couplings is obtained by external
driving. This means that one can realize a whole class of anisotropic
$XY\!Z$ models with ferromagnetic and/or  
anti-ferromagnetic correlations.  
To illustrate the rich physics that can be explored with this system
we discuss the phase diagram of the $1D$ $XY\!Z$ chain in an external
field. This case exhibits ferromagnetic as well as anti-ferromagnetic
phases, a magnetized/polarized phase, a spin-flop and a floating
phase~\cite{eran}. Finite size effects relevant for the trapped case
are examined via exact diagonalization. This reveals the appearance of
a devil's staircase manifested in the form of spin density
waves. Finally, we discuss how to experimentally probe and manipulate
the spin degrees of freedom. 

{\it $p$-orbital Bose system.}-- We consider bosonic atoms of mass
$m$ in a 2D optical lattice of the form  $V(\mathbf{r}) =
V_{x}\sin^2(k_{x}x)+V_y\sin^2(k_yy)$. Assuming that all atoms are in
the first excited bands, 
the tight-binding Hamiltonian is  
\begin{gather}
\hat H=-\sum_{ ij,\alpha}t^\alpha_{ij}\hat a_{i,\alpha}^\dagger\hat
a_{j,\alpha}+\! 
\sum_{i,\alpha}\!\left[\frac{U_{\alpha\alpha}}{2}\hat
  n_{i,\alpha}(\hat n_{i,\alpha}-1)+E_\alpha^p\hat
  n_{i,\alpha}\right]\nonumber\\ 
+\sum_{i,\alpha\neq\alpha'}\left(
U_{\alpha\alpha'}\hat n_{i,\alpha}\hat n_{i,\alpha'}+\frac
{U_{\alpha\alpha'}} 2 \hat a_{i,\alpha}^\dagger\hat
a_{i,\alpha}^\dagger 
\hat a_{i,\alpha'}\hat a_{i,\alpha'}\right).
\label{Hubbard}
\end{gather}
Here $\hat a_{i,\alpha}^\dagger$ creates a bosonic particle in the orbital
$\alpha=p_x,\,p_y$  at site $i$, $\hat n_{i,\alpha}=\hat
a_{i,\alpha}^\dagger\hat a_{i,\alpha}$, and the sum is over nearest
neighbors $i,j$. The tunneling matrix elements are given by
$t^\alpha_{ij} = - \int
d\mathbf{r}\,{w^\alpha_i}(\mathbf{r})^*\left[-\hbar^2\nabla^2/2m
  +V(\mathbf{r})\right]w^\alpha_j (\mathbf{r})$ where
$w^\alpha_i(\mathbf{r})$ is the 
Wannier function of orbital $\alpha$ at site $i$. Note that
$t^\alpha_{ij}$ is anisotropic. For instance, a boson in the
$p_x$-orbital has a much larger tunneling rate in the $x$-direction
than in the $y$-direction. The coupling constants are given by  $U_{\alpha
 \alpha'} = U_0 \int d\mathbf{r}\,
|w^\alpha_i(\mathbf{r})|^2|w^{\alpha'}_i(\mathbf{r})|^2$, with $U_0>0$
the  onsite interaction strength determined by the
scattering length. The last term in (\ref{Hubbard}) is the orbital
changing term describing the flipping of a pair of atoms from the
state $\alpha'$ to the state $\alpha$. Note that this term is absent 
in the case of fermionic atoms. 

 
{\it Effective spin Hamiltonian.}-- We are interested  in the
physics of the Mott insulator phase with unit filling in the strongly
repulsive limit  
$|t^\alpha_{ij}|^2\ll U_{\alpha\alpha'}$. Projecting onto the  
Mott space of singly occupied sites with the operator $\hat P$, the
Schr\"odinger equation becomes $\hat H_{\rm Mott}\hat
P|\psi\rangle=E\hat P|\psi\rangle$ 
with  $\hat H_{\rm Mott}=-\hat P\hat H(\hat H_Q-E)^{-1}\hat H\hat
P$. Here $\hat Q=1-\hat P$  and $\hat H_Q=\hat Q\hat H\hat
Q$~\cite{auerbach}. Since $E\sim t^2/U$, we can take $( \hat
H_Q-E)^{-1}=\hat H_Q^{-1}$.  

The space of doubly occupied states of a given site
$j$ is three-dimensional and spanned by  
$|p_xp_x\rangle=2^{-1/2}\hat a_{jx}^\dagger\hat a_{jx}^\dagger|0\rangle$,
$|p_yp_y\rangle=2^{-1/2}\hat a_{jy}^\dagger\hat a_{jy}^\dagger|0\rangle$,  
and $|p_xp_y\rangle=\hat a_{jx}^\dagger\hat a_{jy}^\dagger|0\rangle$. In
this space, it is straightforward to find $\hat H_Q$ from
(\ref{Hubbard}), and  subsequent
inversion yields
\begin{equation}
\hat H_Q^{-1}=
\begin{pmatrix}
U_{yy}/U^2&-U_{xy}/U^2&0 \\
-U_{xy}/U^2&U_{xx}/U^2&0 \\
0&0&1/2U_{xy} \\
\end{pmatrix}
\label{Hqinv}
\end{equation} 
with $U^2=U_{xx}U_{yy}-U_{xy}^2$. In particular, the off-diagonal
terms in $\hat H_Q^{-1}$ derive from the orbital changing term.
Using~(\ref{Hqinv}) we can now calculate all possible matrix elements
of $\hat H_{Mott}$ in the Mott space,
\begin{gather}
\hat H_{\rm Mott}=-\sum_{ij,\alpha}\left(\frac
     {2|t^\alpha_{ij}|^2U_{\bar\alpha\bar\alpha}} {U^2}\hat
     n_{i,\alpha}\hat n_{j,\alpha}  
+\frac {|t^\alpha_{ij}|^2} {2U_{xy}}\hat n_{i,\alpha}\hat
n_{j,\bar\alpha}\right.\nonumber\\ 
-\left.\frac {2t^x_{ij}t^y_{ji}U_{xy}} {U^2}
\hat a_{i,\alpha}^\dagger\hat a_{i,\bar\alpha}\hat
a_{j,\alpha}^\dagger\hat a_{j,\bar\alpha} 
+\frac {t^x_{ij}t^{y}_{ji}} {2U_{xy}}
\hat a_{i,\alpha}^\dagger\hat a_{i,\bar\alpha}\hat
a_{j,\bar\alpha}^\dagger\hat a_{j,\alpha}\right) 
\label{HMottFinal}
\end{gather}
where $\bar x=y$, and $\bar y=x$. By further employing the Schwinger
angular momentum representation, 
$\hat{S}^z_i = \frac{1}{2}(\fd_{xi}\f_{xi} - \fd_{yi}\f_{yi})$,
$\hat{S}^{+}_i = \hat S^x_i + i\hat S^y_i = \fd_{xi}\f_{yi}$ and $\hat{S}^{-}_i
= \hat S^x_i - i\hat S^y_i = \fd_{yi}\f_{xi}$, together with the constraint
$\fd_{xi}\f_{xi} + 
\fd_{yi}\f_{yi} =1$, we can (ignoring irrelevant constants)
map~(\ref{HMottFinal}) onto a spin-1/2 $XY\!Z$ model in an
external field~\cite{coms}  
\begin{gather}
\hat H_{XY\!Z}=\sum_{\langle ij\rangle}J_{ij}\left[(1+\gamma)\hat S^x_i\hat
  S^x_j+(1-\gamma)\hat S^y_i\hat S^y_j\right]\nonumber\\ 
+\sum_{\langle ij\rangle}\Delta_{ij} \hat S^z_i\hat S^z_j+h\sum_i\hat S_i^z.
\label{HMottSpinFinal}
\end{gather}
Here, $\langle i,j\rangle$ means summing over each nearest neighbor
pair $i,j$ only once.  The couplings are given by
$J_{ij}=-2t^x_{ij}t^y_{ji}/U_{xy}$, $\gamma=-4U_{xy}^2/U^2$, and  
$\Delta_{ij} = -4(|t^x_{ij}|^2U_{yy}+|t^y_{ij}|^2U_{xx})/U^2
+(|t^x_{ij}|^2 +|t^y_{ij}|^2)/U_{xy}$. 
The magnetic field is  $h=4\sum_{\langle
  ij\rangle}(|t_{ij}^y|^2U_{xx}-|t_{ij}^x|^2U_{yy})/U^2+E_{p_x} -
E_{p_y}$, where $E_{\alpha}$ is the onsite energy of the orbital
$\alpha$. 

Equation (\ref{HMottSpinFinal}) is a main result of this paper. It
demonstrates how $p$-orbital  
bosons in a 2D optical lattice can realize the  $XY\!Z$ quantum
spin-$1/2$ Heisenberg model. Several interesting facts should be
noted. First, $t^x_{ij}t^y_{ji}<0$ due to the symmetry 
 of the $p$-orbitals~\cite{girvin} and therefore $J_{ij}>0$. 
Furthermore, since $|\gamma|<1$ we  have  anti-ferromagnetic
 instead of the usual ferromagnetic couplings  for bosons. Also, we
 obtain the $XY\!Z$ model when  $\gamma\neq 0$. 
 The presence of $\gamma$ can be traced  to the orbital changing term
 in Eq.~(\ref{Hubbard}),  
 which reduces the continuous $U(1)$ symmetry of $\hat S^x$ and $\hat
 S^y$ to a set of $Z_2$ 
symmetries. The $Z_2$ symmetries reflect the `parity' conservation in
the original bosonic picture which classifies the
many-body states according to total even or odd number of atoms in the
$p_x$ and $p_y$ orbitals. Since the orbital changing term is absent
for fermions, the $XY\!Z$ model with anisotropic  
coupling is a peculiar feature of bosons in the $p$ band. 
We emphasize that the above derivation makes no assumptions regarding
the geometry of the 2D lattice - i.e.\  it can be square, hexagonal
etc.

{\it 1D $XY\!Z$ phase diagram.}-- To illustrate the rich physics of
the $XY\!Z$ model, we now focus on 
the case of a 1D lattice where where quantum fluctuations are especially
pronounced. Note that by increasing both the lattice  
amplitude and spacing in the $y$ direction keeping $V_yk_y^2\simeq
V_xk_x^2$, one can exponentially suppress tunneling in the
$y$ direction to obtain a 1D model, while the $p_x$ and $p_y$ orbitals
are still quasi-degenerate~\cite{liu1D}. In the 1D setting, we will drop
the "direction" subscript $ij$ on the coupling constants.   

For 1D, the importance of the orbital changing term can be further
illuminated, by employing the Jordan-Wigner
transformation  $\hat{S}_i^-=e^{i\pi\sum_{j=1}^{i-1}\hat{c}_j^\dagger\hat{c}_j}\hat{c}_i$   
for fermionic operators $\hat{c}_i$. The result is the fermionic Hamiltonian 
\begin{gather}
\hat{H}_K/J = \!\sum_n\!\Big[\!\left(\fod_n\fo_{n+1}\! +\!
  \fod_{n+1}\fo_n\right)\! +\! 
 \gamma\left(\fod_n\fod_{n+1}\! +\! \fo_{n+1}\fo_n\right)\! + \nonumber \\
\frac{\Delta}{J}\left(\fod_n\fo_n +
\frac{1}{2}\right)\left(\fod_{n+1}\fo_{n+1} - \frac{1}{2}\right) \;+ 
\frac{h}{J}\left(\fod_n\fo_n - \frac{1}{2}\right)\Big]. 
\label{fermi_ham}    
\end{gather}
We see that $\gamma\neq 0$ leads to a pairing term that typically 
opens a gap in the energy spectrum. Incidentally the limit of $\Delta
\rightarrow 0$ in Eq.~(\ref{fermi_ham}) is a realization of the
Kitaev chain~\cite{kitaev}.   

\begin{figure}
\centerline{\includegraphics[width=0.8\columnwidth]{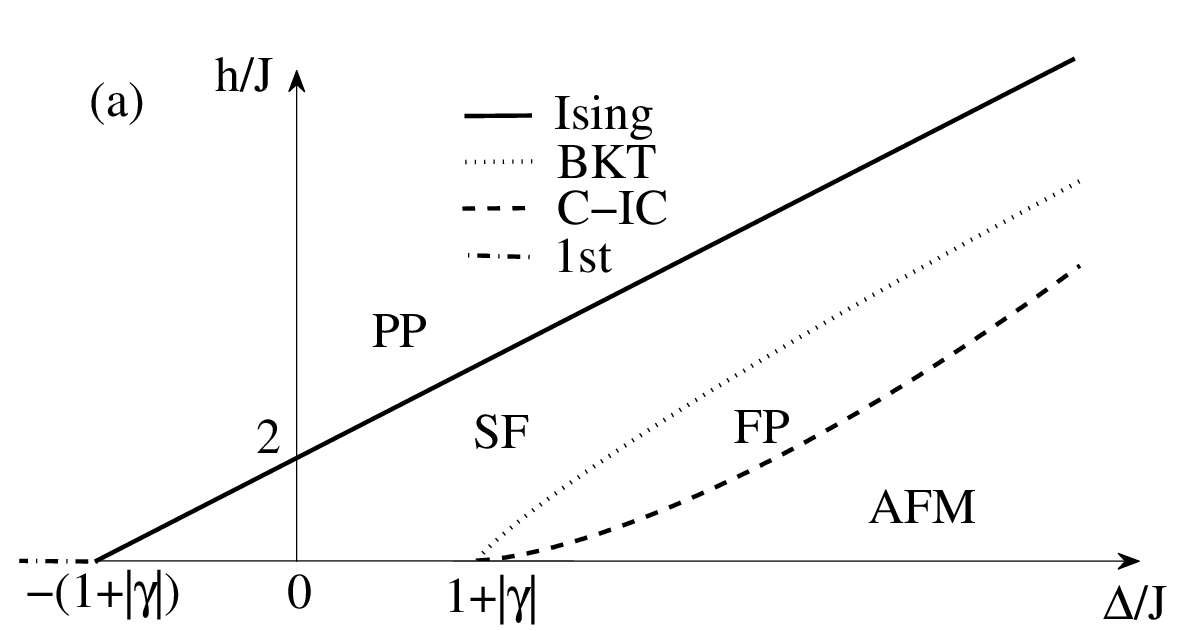}}
\centerline{\includegraphics[width=0.82\columnwidth]{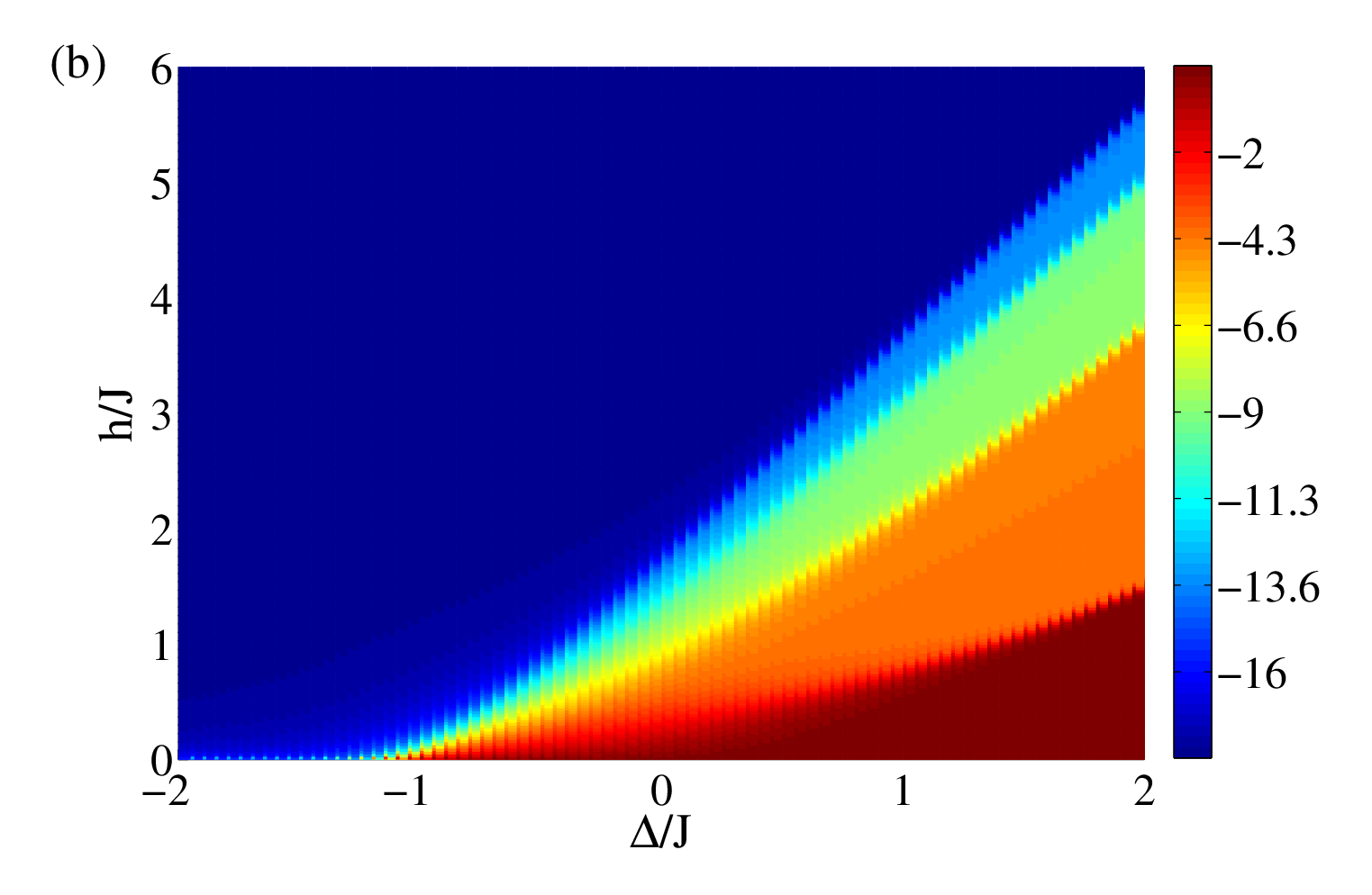}}
\caption{(Color online) (a) Schematic
  phase diagram of the $XY\!Z$ chain. (b) Finite size 'phase diagram'
  obtained by exact diagonalization of 18 spins. The finite size 'phase
  diagram' comprises an incomplete devil's staircase of SDW between the
  PP and AFM phases. The anisotropy parameter is $\gamma=0.2$ in (b).} 
\label{fig1}
\end{figure}
 The schematic phase diagram is illustrated in
Fig.~\ref{fig1} (a).   
At zero field, the $XY\!Z$ model is
integrable~\cite{Takhtadzhan}. For large positive values of $\Delta/J$
the system is anti-ferromagnetic (AFM) in the $z$ direction. Small
values of 
$\Delta/J$ are characterized by N\'{e}el ordering in the
$y$ direction and the system is in the so-called spin-flop phase
(SF). The $h = 0$ line for large negative values of $\Delta/J$ is
characterized by a ferromagnetic phase (FM) in the 
$z$ direction, and for all the cases, the limit of large external
field displays a magnetized phase (PP),
where the spins align along the orientation of the field in the
$z$ direction. These three phases also characterize the phase diagram
of the $XXZ$ model in a longitudinal field~\cite{mikeska}. However,
for non-zero  anisotropy $\gamma$, a gapless floating phase (FP)
emerges between the SF and the AFM phases which is characterized by
power-law decay of the 
correlations~\cite{bak,eran,com}. The transition from the AFM to the 
FP is of the commensurate-incommensurate (C-IC) type whereas 
the transition between the FP and SF phases is of the
Berezinsky-Kosterlitz-Thouless (BKT) type. For $\Delta<-(1+|\gamma|)$
there is a first order transition at $h=0$ between the two polarized
phases. Finally, there is an Ising transition between the PP and the
SF phases.     

The experimental realization of
the Heisenberg model  will inevitably involve finite
size effects due to the harmonic trapping potential. Within the local
density approximation, the trap renormalizes the couplings so 
that they become spatially dependent~\cite{fep1}, but this effect can be
negligible if the orbitals are small compared to the length scale of
the trap. In the regime of strong repulsion, the main effect of the
trap is instead that it gives rise to ``wedding cake'' structures with
Mott regions of integer 
 filling. This effect was observed in the lowest band Bose-Hubbard
model~\cite{maciek}, and predicted theoretically to occur for
anti-ferromagnetic systems~\cite{AndersenBruun}.   
To examine finite size effects, we have performed exact
diagonalization in a chain with 18 spins with open boundary conditions.
Figure~\ref{fig1} (b) displays the resulting finite size 'phase diagram'.
The colors correspond to different values of the total magnetization
$M=\sum_i\langle \hat{S}_i^z\rangle$ of the ground state. While the PP
phase and the AMF phase are both clearly visible, the numerical
results reveal a step like structure of the magnetization in between
the two phases. We attribute these steps in $M$ to  
a devil's staircase structure of spin-density-waves (SDW). 
As we see from Fig.~\ref{fig1} (b), it is only possible to give a
numerical result for the PP-SF Ising transition. In 
particular, the C-IC and BKT transitions are overshadowed by the 
transitions between SDW. In the thermodynamic
limit the staircase becomes complete and the changes in
$M$ become smooth. One then recovers the phase diagram of
Fig.~\ref{fig1} (a). These transitions, between different SDW, are
more pronounced for moderate systems sizes. For a typical
experimental system with $\sim$50 sites, for example, we estimate
$\sim$15 different SDW between the AFM and PP phases.

{\it Measurements and manipulations.}-- While time-of-flight
measurements can reveal some of the 
phases~\cite{complex}, single-site addressing
techniques~\cite{single_site} will be much more powerful when extracting
correlation functions. To address single orbital states or even
perform spin rotations, one may borrow techniques developed for
trapped ions~\cite{leshouches}. Making use of the symmetries of the 
$p_x$ and $p_y$ orbitals,
stimulated Raman transitions can drive both sideband and 
carrier transitions for the chosen orbitals in the Lamb-Dicke
regime. These transitions can be made so short that the system
is essentially frozen during the operation. Driving sideband
transitions in this way, spin rotations   
may be implemented. For example, a spin rotation around $x$ is
achieved by driving the red-sidebands for both
orbitals~\cite{coms}. As a result, the two $p$ orbitals are coupled to
the $s$ orbital in a $V$ configuration and in 
the large detuned case an adiabatic elimination of the $s$ band
gives an effective coupling between the $p_x$ and
$p_y$ orbitals~\cite{bruce}. This scheme, thus, realizes an effective 
spin Hamiltonian
$\hat{H}_x^{(i)}=\frac{\Omega_x\Omega_y}{\Delta_\mathrm{ps}}\hat{S}_i^x$
with $\Omega_\alpha$ the effective Rabi frequencies and
$\Delta_\mathrm{ps}$ the detuning. Alternatively, Stark-shifting one
of the $p$ orbitals results in a rotation around $z$. Since the 
spin operators do not commute, any rotation can be realized from these
two operations. Performing fluorescence on single orbital states by
driving the carrier 
transition acts as measuring $\hat{S}_i^z$. This combined with the
above mentioned rotations makes it possible to measure the spin at any
site in any direction~\cite{coms, leshouches}.  

\begin{figure}
\includegraphics[width=0.8\columnwidth]{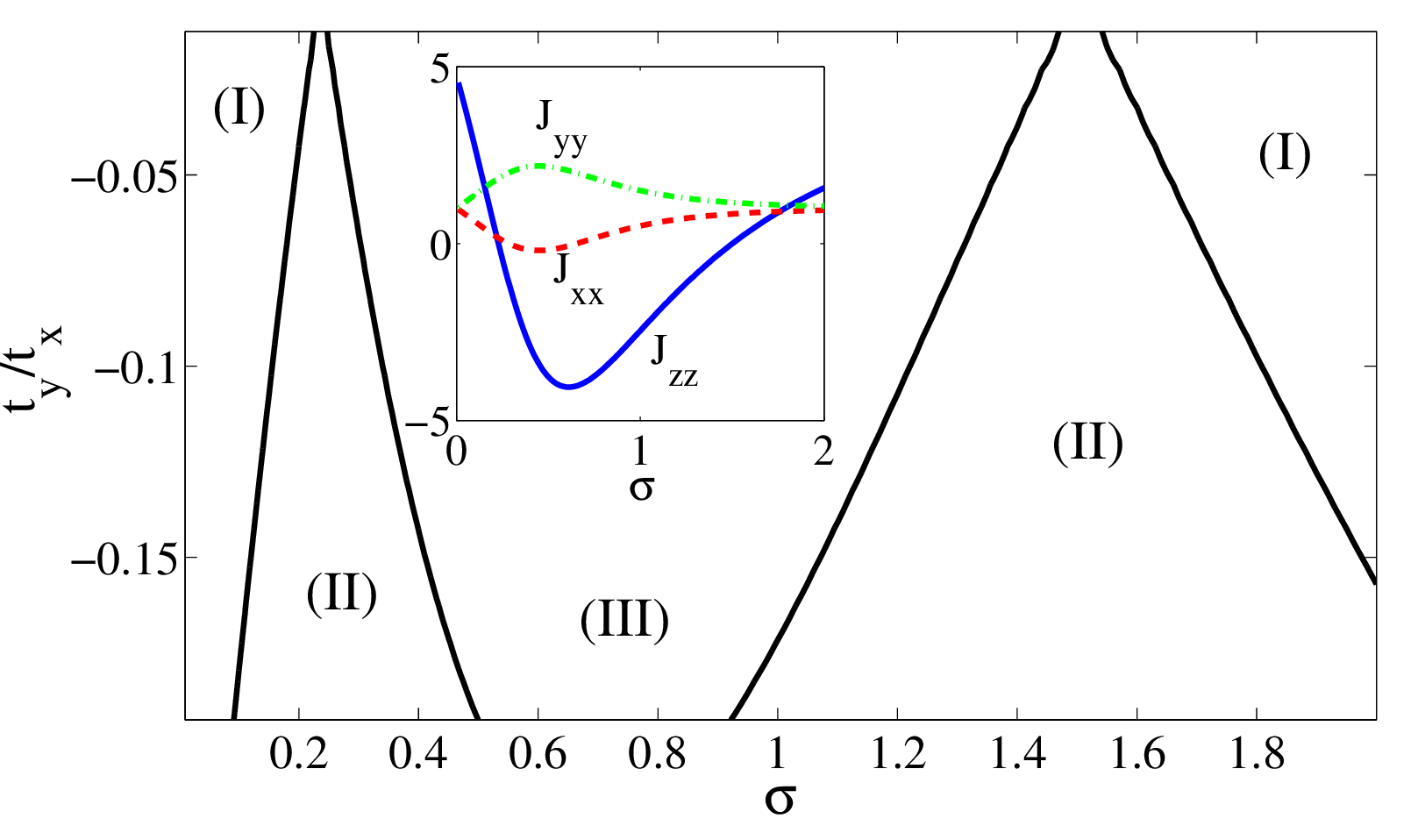}
\caption{(Color online)  Different types of models are achieved by
  varying the relative tunneling strength and the relative orbital
  squeezing. The three different parameter regions are: (I)
  anti-ferromagnetic couplings in all spin components with
  $\Delta>J(1+|\gamma|)$, (II) ferromagnetic or anti-ferromagnetic
  couplings in the $z$-component
  and anti-ferromagnetic in the $y$-component with
  $J(1+|\gamma|)>|\Delta|$, and (III) same as in (II) but with
  $|\Delta|>J(1+|\gamma|)$. The inset shows one example of the spin
  parameters $J_{xx}=(1+\gamma)$, $J_{yy}=(1-\gamma)$, and
  $J_{zz}=\Delta/J$ for $t_y/t_x=-0.1$.} 
\label{fig2}
\end{figure}

{\it Tuning of couplings.}-- For a square optical lattice, we have
$U_{xx}=U_{yy}$. Moreover, in the harmonic approximation
$U_{xy}=U_{xx}/3$, from which it follows that $\Delta<0$ and
$\gamma=-1/2$. This gives ferromagnetic couplings for the 
 $z$ component of neighboring spins, while the interactions between
$x$ and between the $y$ components have anti-ferromagnetic
couplings. We now show how the relative strength and sign of the
different couplings can be controlled by 
squeezing one of the orbital states.  Such squeezing can be
accomplished by again driving the carrier transition of either of the
two orbitals, dispersively with a spatially dependent
field~\cite{coms}. The shape of the drive can be chosen such that the
resulting Stark shift is 
weaker in the center of the sites, resulting in a narrowing of the
orbital. To be specific, assume that the ratio $\sigma$ of the
harmonic length scales of 
the $p_x$ and $p_y$ orbitals in the $y$ direction is tuned. A
straightforward calculation using harmonic oscillator functions yields  
$\alpha\equiv U_{xx}/U_{xy}=2^{-3/2}3(1+\sigma^2)^{3/2}/\sigma$ and 
$\beta\equiv U_{yy}/U_{xy}=2^{-3/2}3(1+\sigma^2)^{3/2}$. The coupling
constants now depend on $\sigma$  as  
$\Delta/J=2t^x(t^y)^{-1}\beta/(\alpha\beta-1)+2t^y(t^x)^{-1}
\alpha/(\alpha\beta-1)-  
(t^x/t^y+t^y/t^x)/2$    
and $\gamma=-4/(\alpha\beta-1)$.    
 The inset in Fig.~\ref{fig2} displays the three coupling parameters
 as a function of $\sigma$ for $|t^x/t^y|=0.1$. We see that the
 relative size and even the sign  
 of the couplings can be tuned by varying $\sigma$. In particular,
 while $\hat S_y$ always has AFM couplings, they can be made both FM
 or AFM for  $\hat S_x$ and $\hat S_z$. 
  In the main part of Fig.\ \ref{fig2}, we sketch the different
  accessible models as a function of $t^y/t^x$ and $\sigma$.  
This clearly demonstrates that one can realize a whole class of
$XY\!Z$ spin chains by using this method.

{\it Experimental realization.}--  In Ref.~\cite{TMuller}, the
experimental realization of $p$-orbital bosons in an effective 1D
optical lattice with a life-time of several milliseconds was reported. 
With an average number of approximately two atoms per
site, the atoms could tunnel hundreds of times in the $p$ band before
decaying. Since the main decay mechanism stems from atom
collisions~\cite{TomaszSpin,girvin},  
an increase of up to a factor of 5 in the lifetime is
expected when there is only one atom per site~\cite{TMuller}.
Typical values of the couplings
can be estimated from the overlap integrals of neighboring Wannier
functions. Considering $^{87}$Rb atoms, 
$\lambda_{lat} = 843\,n$m and $V_x = 30E_R$, $V_y = 50E_R$ and $V_z =
60E_R$, we obtain $J/E_R \sim 0.01$ and the
characteristic tunneling 
time $\tau = \hbar/J\sim 5\,m$s.
This corresponds to a few dozens of times smaller than the expected
lifetimes~\cite{TMuller}, which should allow for experimental
explorations of our results since relaxation typically occurs on a
scale less than ten tuneling times~\cite{GreinerPaper}. In addition,
as pointed out 
in~\cite{coms}, it is possible to increase the lifetimes even further
with the use of external driving.

A major experimental challenge is to achieve a unit filling of the
$p$ band. This could be achieved by having an excess number
of atoms in the $p$ band and then adiabatically opening up the trap
such that the unit filling is reached. A minority of sites will still
be populated, however, 
by immobile $s$-orbital atoms. Since the interaction energy
between  $s$- and $p$-orbital atoms is higher than between two
$p$-orbital atoms, processes involving 
$s$-orbital atoms will be suppressed. The presence of atoms in
the $s$ band corresponds therefore to  
introducing static disorder in the system~\cite{coms}. This may 
affect correlations~\cite{dis}, but the qualitative physics will
remain unchanged for concentrations close to a unit filling. 
A more detailed study of this interesting effect is beyond the scope
of the present work. 

As a final remark we note that the spin correlations
discussed here will emerge at temperatures  $k_BT\lesssim
J\sim t^2/U$~\cite{spin3}. In addition, we estimate the required
entropy~\cite{Kollath} 
by equating the critical temperature $T_c$ to the gap between
the ground and first excited states in the
anti-ferromagnetic phase. Using the energy spectrum obtained from
exact diagonalization, $S = (E - F)/T_c$ yields the entropy per
particle $S/N =
0.06k_B$. Experimentally one has in fact already achieved $S/N =
0.05k_B$~\cite{GreinerTalk}, which
indicates that our results are within experimental reach.



{\it Conclusions.}-- We showed that the Mott regime of unit filling
of bosonic atoms in the first excited bands of a 2D 
optical lattice realizes the spin-$1/2$ $XY\!Z$ quantum Heisenberg
model. We then illustrated the rich physics of this model by examining the
phase diagram of the 1D case. Finite size effects relevant to the
trapped systems were discussed in detail. We
proposed a method to control the strength and relative size of the
spin couplings thereby demonstrating how one can realize a whole class
of $XY\!Z$ models.
We finally discussed experimental issues related to the realization of
this model.  We end by noting, that recent experiments reported a 
$\sim$99\% loading fidelity of bosons into the $d$-band~\cite{d_band},
which indeed opens possibilities to probe rich physics beyond spin-$1/2$
chains. 

{\it Acknowledgments.}-- We thank Alexander Altland, Alessandro De
Martino, Henrik Johannesson, Stephen Powell, Eran Sela, and Tomasz
Sowi\'nski for helpful discussions. We acknowledge financial support
from the Swedish research council (VR).  GMB acknowledges financial
support from NORDITA.



\newpage
\begin{center}
\large
{\bf SUPPLEMENTARY MATERIAL}
\normalsize
\end{center}

\section{Derivation of the effective spin model}
We are interested in the strong coupling regime where the
system is deep in the Mott insulator phase with a unit filling $n=1$
of the lattice sites. A natural way of analyzing this limit
involves the use of projection operators that divide the Hilbert space
of the associated eigenvalue problem in orthogonal subspaces according to
 site occupations. We  define
the $\hat{P}$ and $\hat{Q}$ operators that project, respectively, into
the subspace of states with a unit occupation and into the perpendicular
subspace. They decompose the 
eigenvalue equation $\hat{H}|\Psi\rangle = E|\Psi\rangle$, with $E$
its associated energy, in the form    
\begin{equation}\label{eig_problem}
\begin{array}{l}
\left(\hat{Q}\hat{H}_t\hat{P}\! +\! \hat{Q}\hat{H}_t\hat{Q}\! +\!
  \hat{Q}\hat{H}_U\hat{P}\! +\! \hat{Q}\hat{H}_U\hat{Q}\right)|\Psi\rangle =
  E\hat{Q}|\Psi\rangle\\ \\
\left(\hat{P}\hat{H}_t\hat{P}\! +\! \hat{P}\hat{H}_t\hat{Q}\! +\!
  \hat{P}\hat{H}_U\hat{P}\! +\! \hat{P}\hat{H}_U\hat{Q}\right)|\Psi\rangle =
  E\hat{P}|\Psi\rangle, 
\end{array}
\end{equation}
where $\hat{H}_U$ is the interaction part of the Hamiltonian. Since
$\hat{Q}\hat{H}_t\hat{Q}$, $\hat{Q}\hat{H}_U\hat{P}$,
$\hat{P}\hat{H}_U\hat{P}$, and  
$\hat{P}\hat{H}_t\hat{P}$  all vanish,  it follows that 
\begin{equation}\label{qpsi}
\hat{Q}|\Psi\rangle = -\frac{1}{\hat{Q}\hat{H}\hat{Q} -
  E}\hat{Q}\hat{H}_t\hat{P}|\Psi\rangle. 
\end{equation}
By further substitution of Eq.~(\ref{qpsi}) in the eigenvalue
equation, we are left with the Hamiltonian which describes the one
particle Mott phase of $p$-orbital bosons
\begin{equation}\label{Mott1}
\hat{H}_{Mott} =
-\hat{P}\hat{H}_t\hat{Q}\frac{1}{\hat{Q}\hat{H}_U\hat{Q} -
  E}\hat{Q}\hat{H}_t\hat{P}. 
\end{equation}
So far this result is exact. It explicitly shows the role of the
tunneling in the system, namely of coupling the subspace of states
where the sites have unitary occupation with the states that have  one
site doubly 
occupied. First, a particle tunnels, say, from the site $\bj$ to ${\bf 
  j + 1}$, where it interacts with another particle according to what
is described by $\hat{H}_U$. After interaction, one of the
particles is brought back to the site $\bj$, and the final state is
again characterized by lattice sites with a unit filling.

Equation~(\ref{Mott1}) is the starting point in the
derivation of the effective Hamiltonian describing the $n=1$ Mott
phase of $p$-orbital bosons. The procedure is developed here for an
effective 1D system with dynamics along the $x$-axis, but
generalization to the 2D lattice is straightforward. Realization of
the 1D configuration relies on the adjustment of the lattice
parameters, that should contain potential wells much deeper in the
$y$ than in the $x$ axis, but in such a way that the quasi degeneracy
between the different orbital states is still maintained. This means
that $|t_{xy}|, |t_{yy}|\rightarrow 0$, and furthermore, due to the strong
coupling regime condition, we also have that $U_{\alpha\beta}
\gg |t_{xx}|, |t_{yx}|$, $\alpha,\beta = \{x,y\}$. 

Under these assumptions, the operator $1/(\hat{Q}\hat{H}\hat{Q} - E)$
in Eq.~(\ref{Mott1}) can be expanded to second order in
$t/U_{\alpha\beta}$ ($\alpha,\beta=\{x,y\}$) in analogy to the
customary procedure used 
for the Hubbard model at half filling~\cite{Korepin}. In the
tight-binding regime considered here, it is enough to consider the
2-site problem. The basis spanning the  subspace of states with unit
filling  is     
\[
\mathcal{H}_P =\{\vert x,x\rangle\, \vert x,y\rangle, \vert y,
x\rangle, \vert y,y\rangle\},
\]
where $|\alpha,\beta\rangle$ represents the state with an
$p_{\alpha}$-orbital atom in site $i$ and a $p_{\beta}$-orbital atom in site
$j$. The relevant states for the doubly occupied sites is 
\[
\mathcal{H}_Q = \{\vert 0,2x\rangle, \vert 0,
2y\rangle, \vert 0, xy\rangle \}, 
\] 
which span the intermediate states of the projection operation. We
notice, however, that due to the possibility of 
transferring population between the different orbital states, the
projection of the Hamiltonian in the 
$\mathcal{H}_Q$ subspace is not diagonal in this basis of intermediate
states. This is a peculiarity of the present model and derives
entirely from the orbital changing collisions. As a consequence, we
compute $(\hat{H}_Q - E)^{-1}$, with $\hat{H}_Q = 
\hat{Q}\hat{H}\hat{Q}$ by calculating the projected Hamiltonian in the
$\mathcal{H}_Q$ subspace and taking its corresponding inverse. Since
$E\sim t^2/U_{\alpha\beta}$,
it is justified to ignore $E$ and to consider $(\hat{H}_Q - E)^{-1} \approx
\hat{H}_Q^{-1}$. Explicitly,  
\[
\hat{H}_Q = \left(\begin{array}{ccc}
U_{xx} & U_{xy} & 0\\
U_{xy} & U_{yy} & 0\\
0 & 0 & 2U_{xy}\end{array}\right)
\]
giving 
\[\hat{H}_Q^{-1} = \left(\begin{array}{ccc}
U_{yy}/U^2 & -U_{xy}/U^2 & 0\\
-U_{xy}/U^2 & U_{xx}/U^2 & 0\\
0 & 0 & U_{xy}/2\end{array}\right),
\]
where $U^2 \equiv U_{xx}U_{yy} -
U_{xy}^2$.  

We determine the final form of the effective Hamiltonian by computing
the relevant matrix elements of~(\ref{Mott1}). To this end, we
consider in detail all the different cases where the resulting action
of the operator $\hat{H}_{Mott}$ of Eq.~(\ref{Mott1}) in the states of
the $\mathcal{H}_P$ 
subspace yield non vanishing contribution.

From states of the type $\vert\alpha_i,\alpha_j\rangle$
\[
\fd_{\alpha, i}\f_{\alpha, j}\hat{H}_Q^{-1}\fd_{\alpha, j}\f_{\alpha,
  i}\vert\alpha_i,\alpha_j\rangle = \fd_{\alpha, i}\f_{\alpha,
  j}\hat{H}^{-1}_Q\sqrt{2}\vert 0, 2\alpha_j\rangle
\]
\[
=\! \sqrt{2}\fd_{\alpha, i}\f_{\alpha,
  j}\!\!\left(\frac{U_{\beta\beta}}{U^2}\vert 0, 2\alpha_j\rangle -
\frac{U_{\alpha\beta}}{U^2}\vert 0, 2\beta_j\rangle\!\right)\! =\!\!
\frac{2U_{\beta\beta}}{U^2}\vert\alpha_i,\alpha_j\rangle 
\]
the effective Hamiltonian acquires a term of the form 
\[
-\sum_{\langle
  i,j\rangle}\sum_{\alpha}\frac{2|t^{\alpha}_{ij}|^2U_{\beta\beta}}{U^2}\hat{n}_{\alpha,
  i}\hat{n}_{\alpha, j}. 
\]
In these and the following expressions, it is understood that
$\beta\neq\alpha$. In the same way, from the states of the type
$\vert\alpha_i,\beta_j\rangle$, 
\[
\fd_{\alpha, i}\f_{\alpha, j}\hat{H}_Q^{-1}\fd_{\alpha, j}\f_{\alpha,
  i}\vert\alpha_i,\beta_j\rangle = \fd_{\alpha, i}\f_{\alpha,
  j}\hat{H}^{-1}_Q\vert 0,\alpha_j\beta_j\rangle
\]
\[
= \frac{1}{2U_{xy}}\fd_{\alpha, i}\f_{\alpha,
  j}\vert 0, \alpha_j\beta_j\rangle =
\frac{1}{2U_{xy}}\vert\alpha_i,\beta_j\rangle, 
\]
corresponding to the operator
\[
-\sum_{\langle i,j\rangle}\sum_{\alpha}\frac{|t^{\alpha}_{ij}|^2}{2U_{xy}}\hat{n}_{\alpha,
i}\hat{n}_{\beta, j}.
\]
From the states of the type $\vert\beta_i, \alpha_j\rangle$ and the
following process 
\[
\fd_{\alpha, i}\f_{\alpha, j}\hat{H}_Q^{-1}\fd_{\beta, j}\f_{\beta,
  i}\vert\beta_i,\alpha_j\rangle = \fd_{\alpha, i}\f_{\alpha,
  j}\hat{H}^{-1}_Q\vert 0,\alpha_j\beta_j\rangle
\]
\[
= \frac{1}{2U_{xy}}\fd_{\alpha, i}\f_{\alpha,
  j}\vert 0, \alpha_j\beta_j\rangle =
\frac{1}{2U_{xy}}\vert\alpha_i,\beta_j\rangle, 
\]
the Hamiltonian gains a contribution as 
\[
-\sum_{\langle i,j\rangle}\sum_{\alpha}\frac{t^{\alpha}_{ji}t^{\beta}_{ij} }{2U_{xy}}
\hat a^{\dagger}_{\alpha, i}\hat a_{\beta, i}\hat a^{\dagger}_{\beta, j}\hat a_{\alpha, j}
\]
Finally, we consider the states of the type $\vert\beta_i,
\beta_j\rangle$, 
 \[
\fd_{\alpha, i}\f_{\alpha, j}\hat{H}_Q^{-1}\fd_{\beta, j}\f_{\beta,
  i}\vert\beta_i,\beta_j\rangle = \fd_{\alpha, i}\f_{\alpha,
  j}\hat{H}^{-1}_Q\sqrt{2}\vert 0,2\beta_j\rangle
\]
\[
=\! \sqrt{2}\fd_{\alpha, i}\f_{\alpha,
  j}\!\!\left(\!\frac{U_{\alpha\alpha}}{U^2}\vert 0, 2\beta_j\rangle\! -
\!\frac{U_{xy}}{U^2}\vert 0, 2\alpha_j\rangle\!\right)\! =\!
-\frac{2U_{xy}}{U^2}\vert\alpha_i, \alpha_j\rangle,
\]
that contribute to the effective Hamiltonian with a term that 
changes the orbital states of the atoms in both sites
\[
\sum_{\langle i,j\rangle}\sum_{\alpha, \alpha\neq\beta}
\frac{2t^{\alpha}_{ji}t^{\beta}_{ij}U_{xy}}{U^2}\fd_{\alpha, i}\f_{\beta, i}\fd_{\alpha,
  j}\f_{\beta, j}.
\]
The resulting expression for the effective Hamiltonian corresponds
thus to   
\begin{equation}\label{Mott1_2}
\begin{array}{l}
\hat{H}_{Mott}= -\displaystyle{\sum_{\langle i,j\rangle}\sum_{\alpha}\left(
\frac{2|t_{\alpha}|^2U_{\beta\beta}}{U^2}\hat{n}_{\alpha,
  i}\hat{n}_{\alpha, j} +
\frac{|t_{\alpha}|^2}{2U_{xy}}\hat{n}_{\alpha, i}\hat{n}_{\beta, j}\right.}
\\ \\
-\displaystyle{\left.\frac{2t_xt_yU_{xy}}{U^2}\fd_{\alpha,
    i}\f_{\beta, i}\fd_{\alpha, j}\f_{\beta, j} +
  \frac{t_xt_y}{2U_{xy}}\fd_{\alpha, i}\f_{\beta, i}\fd_{\beta,
    j}\f_{\alpha, j}\right).} 
\end{array}
\end{equation}

We now use the orbital states to define the Schwinger spin operators  
\begin{equation}
\begin{array}{ccc}
\s^z &= &\displaystyle{\frac{1}{2}(\fd_x\f_x - \fd_y\f_y)}\\ \\
\s^{+}&= &\s^x + i\s^y = \fd_x\f_y\\ \\
\s^{-} &= &\s^x -i\s^y = \fd_y\f_x, 
\end{array}
\end{equation} 
and together with the constraint of unit occupation of the lattice 
sites in the $n=1$ Mott phase, i.e. $\hat{n}_{x, i} + \hat{n}_{y, i} =
1$, we rewrite Eq.~(\ref{Mott1_2}) as
\[
\hat{H}_{Mott} =\!-\!\!\!\sum_{\langle i,j\rangle}\!\!\left(\!J^{zz}\s^z_i\s^z_j\!
+\! J^{xx}\s^x_i\s^x_j\! + \!J^{yy}\s^y_i\s^y_j\!\right) -\sum_i
J^z\s^z_i.  
\]
Thus, within the strong coupling regime, the physics of the $n=1$ Mott
insulator phase is equivalent to  the spin-1/2 Heisenberg $XY\!Z$
model in an external field. In terms of the lattice parameters, the
expressions for the various couplings follow  
\[
J^{xx} = 2\frac{t_xt_y}{U_{xy}}(1 - 4\frac{U_{xy}^2}{U^2})
\]
\[
J^{yy} = 2\frac{t_xt_y}{U_{xy}}(1 + 4\frac{U_{xy}^2}{U^2})
\]
\[
J^{zz} = 4\frac{|t_x|^2U_{yy}}{U^2} + 4\frac{|t_y|^2U_{xx}}{U^2} -
\frac{|t_x|^2}{U_{xy}} - \frac{|t_y|^2}{U_{xy}}
\]
\[
J^{z} = \frac{4|t_x|^2U_{yy}}{U^2} - \frac{4|t_y|^2U_{xx}}{U^2} +
(E^{os}_x - E^{os}_y)
\]

In terms of Eq.~(3) of the main text, we can identify $\Delta=-J^{zz}$,
$h=-J^z$, and $\gamma=-4U_{xy}^2/U^2$. 

\section{Single site addressing of orbital states}
Single site addressing for the present setup implies selective
detection/manipulation of the two orbitals. Since the spin is encoded
in external spatial degrees of freedom rather than internal atomic
electronic states, methods like those described in Refs.~\cite{ion1}
would not work. To control the spatial state 
of the atoms at single sites we may instead apply methods borrowed from trapped ion
physics~\cite{leshouches}. Similar methods were already employed in
the experiment~\cite{TMuller} in order to load bosons from the
$s$ band to the $p$ band. M\"uller {\it et al.} of Ref.~\cite{TMuller}
did not, however, consider single site addressing and more importantly
they did not discuss control of the orbital degree of freedom.  
   
\begin{figure}
\centerline{\includegraphics[width=5cm]{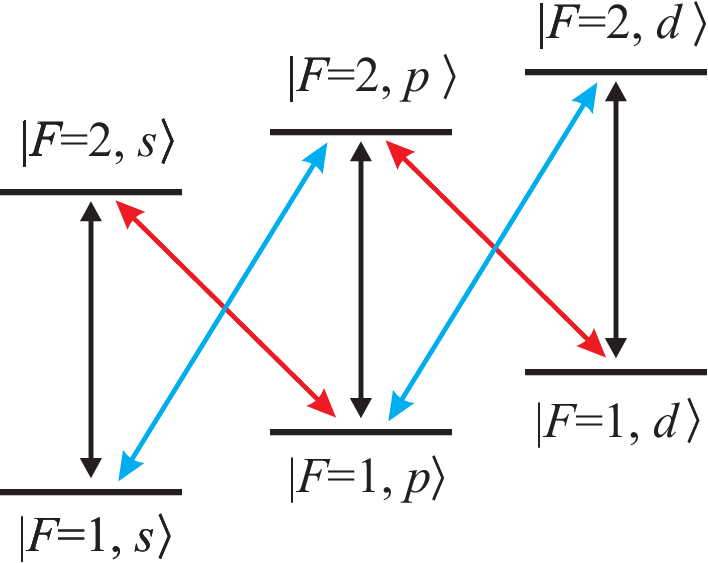}}
\caption{(Color online) Schematic figure of coupling between different
  onsite orbital states. The carrier transition acts upon the internal
  atomic electronic states, while the red and blue sideband
  transitions in addition lower and raise the external vibrational
  state with a single phonon respectively, i.e. couple different
  orbital degree of freedom. } 
\label{fig2}
\end{figure}

Two internal atomic electronic states, e.g. an $F=1$ and an $F=2$
state for $^{87}$Rb atoms, are Raman coupled with two lasers. This
transition is described by the matrix element $\Omega_1\Omega_2\langle
F=2|e^{i(\mathbf{k}_1-\mathbf{k}_2)\cdot\mathbf{x}}|F=1\rangle/\delta$
where $\Omega_i$ and $\mathbf{k}_i$ are the laser amplitudes and wave
vectors, respectively, and $\delta$ the detuning of the transitions
relative to the ancilla electronic state. The spatial dependence of
the lasers will induce couplings between vibrational states of the
atom, i.e. different bands. The time duration for a $\pi/2$-pulse, for
example, can be made very short by making the effective Rabi frequency
$\Omega=\Omega_1\Omega_2/\delta$ large. In particular, this time can
be made short on any other time scale in the system and one can
approximately consider the system dynamics frozen during the applied
pulse. Indeed, the same assumption applies to any single site
addressing in optical lattices. Furthermore, by driving resonant
two-photon transitions we do not need to worry about accidental
degeneracies between other undesired states.  

Deep in the Mott insulator phase, as considered in this work, we can
approximate single sites with two dimensional harmonic oscillators
with frequencies $\omega_\alpha=\sqrt{2V_\alpha k_\alpha^2/m}$. The
Lamb-Dicke parameters~\cite{leshouches,jonas} become
$\eta_\alpha=k_\alpha\sqrt{\hbar/2m\omega_\alpha}$, and within the 
Lamb-Dicke regime ($\eta_\alpha\ll1$) we can neglect multi-phonon
transitions. Thus, in one dimension we have three possible transitions:
($i$) {\it Carrier transitions} - with no change in the vibrational state, ($ii$)
{\it red sideband transitions} - which lower the vibrational state with one
quantum, and ($iii$) {\it blue sideband transitions} - which raise the
vibrational state with one quantum. The various possibilities are
demonstrated in Fig.~\ref{fig2}. 

Since the different transitions are
not degenerate, it is possible to select single transitions by
carefully choosing the frequencies of the lasers. Moreover, considering
for example $\mathbf{k}_1-\mathbf{k}_2=k_x$, i.e. no component in the
$y$ direction, it is possible to only address the $p_x$-orbital. Thus,
we have a method to singly address the different orbitals. 
Full control is achieved when every unitary
$\hat{R}_\beta(\varphi)=e^{-i\hat{S}^\beta \varphi}$, where
$\beta=x,\,y,\,z$ and $\varphi$ is an effective rotation angle, can be
realized. To start with the simplest example, implementation of
$\hat{R}_z(\varphi)$, we first note that since we are considering the
case with a single atom per site $\hat{S}^z=\hat{S}^+\hat{S}^--1$ such
that it is enough to realize the operation of
$\hat{S}^+\hat{S}^-$. This is nothing but a phase shift 
of one of the orbitals. This is most easily done by driving the
carrier transition off-resonantly for one of the two orbitals. Since
the driving is largely detuned it only results in a Stark shift of the
orbital. 

The $\hat{R}_x(\varphi)$ operation is preferably achieved by
simultaneously driving off-resonantly the red sidebands of the two
orbitals. The $s$-band will never get populated due to the large
detuning while instead the transition between the two orbitals can be
made resonant. More precisely, for the three involved states $\{|x,0,0\rangle,\,|0,y,0\rangle,\,|0,0,s\rangle\}$ (with the last entry in the ket-vector being the $s$-orbital) the resulting coupling Hamiltonian in the {\it rotating wave approximation} has the form a $V$-coupled system~\cite{bruce}
\begin{equation}
\hat{H}_{V}=\left[
\begin{array}{ccc}
0 & 0 & \Omega_1\\
0 & 0 & \Omega_2\\
\Omega_1 & \Omega_2 & \delta
\end{array}\right],
\end{equation}
where $\Omega_1$ and $\Omega_2$ have been taken real and for now spatially independent. For $\delta\gg\Omega_1,\,\Omega_2$ we adiabatically eliminate the state $|0,0,s\rangle$ to obtain the desired Hamiltonian generating the rotation $\hat{R}_x(\phi)$, namely
\begin{equation}
\hat{H}_x=\left[\begin{array}{cc} 0 & \Omega \\ \Omega & 0\end{array}\right]=\Omega\hat{S}_x.
\end{equation}

Note that if the Raman transition between the two
orbitals is not resonant, such an action performs a combination of an
$x$- and $z$-rotation. To perform $y$-rotations, one could either
adjust the phases of the lasers or simply note that
$\hat{R}_y(\varphi)=\hat{R}_z(\pi/4)\hat{R}_x(\varphi)\hat{R}_z(-\pi/4)$. With 
this at hand, any manipulation of single site spins can be performed.  
To measure the spin state in a given direction one should combine the
rotations with single site resolved fluorescence (i.e. measuring
$\hat{S}_i^z$)~\cite{haroche}. More precisely, since the drive laser can couple to
the two orbitals individually, one orbital will be transparent to the
laser while the other one will show fluorescence. In other words, one
measures $\hat{S}^z$ on a single site. Other directions of the spin
are measured in the same way, but after the correct rotation has been
implemented to it. Furthermore, with the help of coincident
detection it is possible to also extract correlators
$\langle\hat{S}_i^\alpha\hat{S}_j^\beta\rangle$~\cite{blochnew}. Since there is a
single atom at every site, the ``parity problem''~\cite{ion1} of these techniques
deriving from photon induced atom-atom 
collisions is avoided and thereby loss of atoms will not limit our
measurement procedure. This summarizes how preparation, manipulations,
and detection of single site spins can be performed. 

Finally we note that the methods discussed above can be used in a
broader context. For example, there is a transition between two 
$p$-orbital atoms (one $p_x$- and one $p_y$-orbital atom) and one $s$-
and the $d_{xy}$-orbital atom~\cite{joana}. This transition is
resonant for any parameters $V_x$ and $V_y$ and could in principle
cause rapid decay of the $p$-band state, or even Rabi-type
oscillations between the bands. We note, however, that in the experiment of
Ref.~\cite{TMuller} the collisional decay mechanism was surprisingly
small despite this resonant transition. Nevertheless, one could suppress this resonant transition to increase the life-time even further with
the technique described above: By driving the red sideband for the two
$p$-orbital states dispersively, the $s$ and $p$ bands will be
repelled and thereby this breaks the resonance condition for
$p_x+p_y\rightarrow s+d_{xy}$.     

\section{External parameter control}
The ideas of the previous section can also be utilized to
change the system parameters. The simplest
example is the application of $\hat{S}^+\hat{S}^-$ which implements a 
shift in the external field $h$. Apart from the external field, it is
also desirable to control the coupling in the $z$ component of the
spin, $\Delta$, and especially to tune it from ferromagnetic into
anti-ferromagnetic.  

Using the fact that $|t_x|\gg|t_y|$ we have
\begin{equation}
\Delta\approx-|t_x|^2\left(4\frac{U_{yy}}{U_{xx}U_{yy}-U_{xy}^2}-
\frac{1}{U_{xy}}\right).      
\end{equation}
This is most easily estimated in the harmonic approximation.
Introducing the widths $\sigma_\alpha$ of the orbital wave functions
for the spatial directions $\alpha=x,\,y,\,z$, in this limit  
\begin{equation}\label{rel} 
U_{xx}=U_{yy}=3U_{xy}\equiv\frac{u_0}{\sigma_x\sigma_y\sigma_z},
\end{equation}
where $u_0$ is an effective interaction strength (proportional to the
$s$-wave scattering length). We notice that even though the use of
lattice Wannier functions yields a different ratio between
$U_{\alpha\alpha}$ and $U_{\alpha\beta}$ from what is obtained in the
harmonic limit~\cite{tomasz}, it does not affect the qualitative picture
of the results discussed here. Using (\ref{rel}) in the expression for 
$\Delta$ we find 
\begin{equation}
\Delta=-|t_x|^2\frac{3\sigma_x\sigma_y\sigma_z}{2u_0}<0,
\end{equation}
which yields ferromagnetic couplings for the $z$-component of the spin 
in the harmonic approximation. This is also the case for $\Delta$
computed with numerically obtained Wannier functions for physically
relevant parameters, i.e. within the tight-binding and single-band
approximations and deep in the insulating phase.  

The anti-ferromagnetic regime can be reached, however, again with
techniques of trapped ion physics. 
Instead of changing $h$ by a constant amount in
all sites we consider a Stark shift of one of the two orbitals that is
spatially dependent. This is nothing but a potential that reshapes the
lattice sites differently for the two orbitals. In particular we can
imagine squeezing of one orbital in the $y$ direction. Thus, the two
Wannier functions $w_x(\vec{r})$ and $w_y(\vec{r})$ have the same
widths $\sigma_x$ but different widths $\sigma_y$. This would require
driving the carrier transition with a 
field that has a spatially varying (on the length scale of the $y$
lattice spacing) mode profile. The squeezing in the $y$-direction of
one orbital wave function will not affect the tunneling rates $t_x$
and $t_y$, but change both $U_{yy}$ and $U_{xy}$. We have numerically
verified that by sufficiently strong squeezing, $\Delta$ becomes
negative resulting in anti-ferromagnetic $z$-coupling (see Fig. 2 of
the main text). The 
anti-ferromagnetic coupling can also be obtained be stretching one of
the orbitals in the $y$ direction. We note that this manipulation also
affects the anisotropy parameter $\gamma$ and therefore slightly
shifts the phase boundaries of the phase diagram. However, the
qualitative structure is not changed. As a summary, both $h$ and
$\Delta$ can be controlled solely by external driving, i.e. without
changing the lattice parameters.   

\section{Effective model including imperfections due to $s$-orbital atoms}
Transferring every atom from the $s$ band to the $p$ bands is
experimentally challenging. Even though the possibility of loading
atoms from the lowest band to the $d$ band with 97-99\% fidelity was recently
reported~\cite{d_band}, in experiments involving the $p$ band approximately
20\% of the atoms remain on the lowest band~\cite{TMuller,exp}. In
the experiment reported in Ref.~\cite{TMuller}, the loading resulted
in approximately two $p$-orbital atoms per site. Increasing the
lattice amplitude and 
opening up the trap adiabatically will create an insulating state with
unit filling. The 
$s$-orbital atoms can be considered immobile since the lattice
amplitude will typically be around 20 recoil energies. Thus, random
sites in the lattice will be populated by $s$-orbital
atoms. Energetically it costs more energy to doubly occupy these states
with one $s$- and one $p$-orbital atom than those with two $p$-orbital
atoms, i.e. $U_\mathrm{ps}>U_{\alpha\beta}$ where 
\begin{equation}
U_\mathrm{ps}=U_0\int\,d\mathbf{r}\,|w_i^\alpha(\mathbf{r})|^2|w_i^s(\mathbf{r})|^2
\end{equation}
and $w_i^s(\mathbf{r})$ is the $s$-orbital Wannier function at site $i$. 

\begin{figure}
\centerline{\includegraphics[width=7cm]{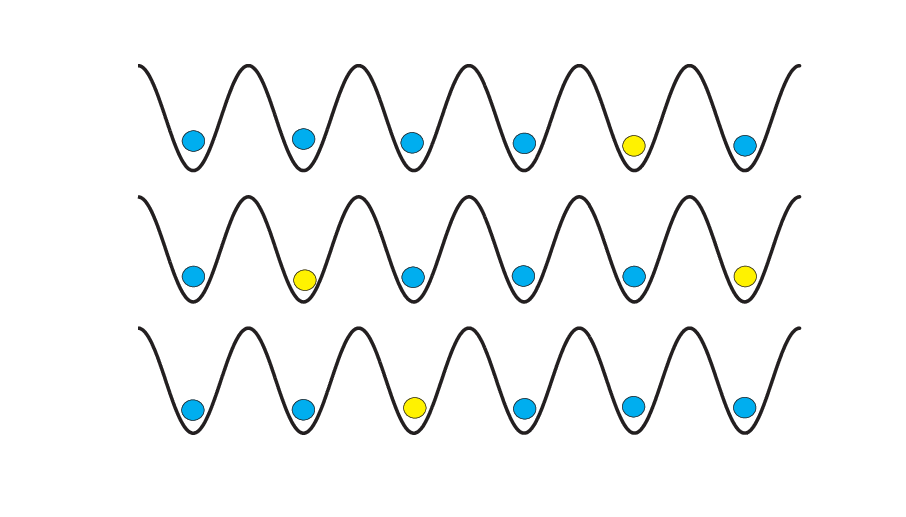}}
\caption{(Color online) Schematic plot of three random experimental
  realization of the insulating state; yellow balls represent
  $s$-orbital atoms and blue ones $p$-orbital atoms. }  
\label{fig3}
\end{figure}

Repeated experimental realizations will prepare different random
configurations as  illustrated in Fig.~\ref{fig3}. The various
configurations are presumably equally probable. If a single realization is
not determined from any measurement, the state will be a statistical
average over all possible configurations. That is, we integrate out
the degrees of freedom of the $s$-band atoms (i.e. average over all
possible configurations constrained to a fixed ratio of $s$-orbital
atoms).  

Let us consider two neighboring sites $i$ and $j$, one with a
$p$-orbital atom and one with an $s$-orbital atom. Since we have
neglected tunneling of $s$-orbital atoms, the only non-vanishing terms
within second order perturbation theory are 
\begin{equation}
\displaystyle{-\frac{t_\alpha^2}{U_\mathrm{ps}}\hat{a}_{\alpha,i}^\dagger
  \hat{a}_{s,j}^\dagger\hat{a}_{\alpha,i}\hat{a}_{s,j}=  
  -\frac{t_\alpha^2}{U_\mathrm{ps}}\hat{n}_{\alpha,i},} 
\end{equation}
where $\alpha=(x,\,y)$, $\hat{a}_{s,j}$ is the annihilation operator
for an $s$-orbital atom at site $j$ and we have used the fact that
$n_{s,j}=1$. Now, since $t_x\neq t_y$ it follows that the presence of
an $s$-orbital shifts the external field $h=J_z$ locally. Thus, the
presence of $s$-orbital atoms will be manifest in local fluctuations
in the external field, i.e. we obtain an $XYZ$ chain with disorder.
\[
\hat{H}_{Mott}^{(\mathrm{dis})} = -\!\!\sum_{\langle
  i,j\rangle}\!\!\!\left(\!J^{zz}\s^z_i\s^z_j\! 
+\! J^{xx}\s^x_i\s^x_j\! + \!J^{yy}\s^y_i\s^y_j\!\right) -\sum_i
J_i^z\s^z_i.  
\]
For few atoms on the lowest band, this effect should not qualitatively
change the results presented in this paper. We expect then
that the disorder is irrelevant~\cite{dis}. For a larger fraction of
$s$-orbital atoms one could expect the disorder to become relevant and
localized phases to appear~\cite{dis}. This interesting topic is,
however, outside the scope of the present paper.

\end{document}